\documentclass[aps,prd,twocolumn,groupedaddress,showpacs]{revtex4}
\usepackage{amssymb}

\usepackage{float,graphicx}

\begin{document}

\title{The Cosmology of $f(R)$ Gravity in the Metric Variational Approach}
\author{Baojiu~Li }
\email[Email address: ]{B.Li@damtp.cam.ac.uk}
\affiliation{Department of Applied Mathematics and Theoretical Physics, Centre for
Mathematical Sciences, Wilberforce Road, University of Cambridge, Cambridge
CB3 0WA, United Kingdom}
\author{John~D.~Barrow}
\email[Email address: ]{J.D.Barrow@damtp.cam.ac.uk}
\affiliation{Department of Applied Mathematics and Theoretical Physics, Centre for
Mathematical Sciences, Wilberforce Road, University of Cambridge, Cambridge
CB3 0WA, United Kingdom}
\date{\today}

\begin{abstract}
We consider the cosmologies that arise in a subclass of $f(R)$
gravity with $f(R)=R+\mu ^{2n+2}/(-R)^{n}$ and $n\in(-1, 0)$ in
the metric (as opposed to the Palatini) variational approach to
deriving the gravitational field equations. The calculations of
the isotropic and homogeneous cosmological models are undertaken
in the Jordan frame and at both the background and the
perturbation levels. For the former, we also discuss the
connection to the Einstein frame in which the extra degree of
freedom in the theory is associated with a scalar field sharing
some of the properties of a 'chameleon' field. For the latter, we
derive the cosmological perturbation equations in general theories
of $f(R)$ gravity in covariant form and implement them numerically
to calculate the cosmic-microwave-background temperature and
matter-power spectra of the cosmological model. The CMB power is
shown to reduce at low $l$'s, and the matter power spectrum is
almost scale-independent at small scales, thus having a similar
shape to that in standard general relativity. These are in stark
contrast with what was found in the Palatini $f(R)$ gravity, where
the CMB power is largely amplified at low $l$'s and the matter
spectrum is strongly scale-dependent at small scales. These
features make the present model more adaptable than that arising
from the Palatini $f(R)$ field equations, and none of the data on
background evolution, CMB power spectrum, or matter power spectrum
currently rule it out.
\end{abstract}

\maketitle












\section{Introduction}

\label{sect:Intro}

Theories of $f(R)$ gravity as an explanation of the dark energy
problem have attracted much recent attention. These theories are
of particular interest since $f(R)$ modifications to general
relativity (GR) appear naturally in the low-energy effective
actions of the quantum gravity and the quantization of fields in
curved spacetime. These theories are also conformally related to
GR with a self-interacting scalar field \cite{BCot} and both the
early time inflation and the late-time acceleration of the
universe could be resulted by a single mechanism in such theories.
In refs.~\cite{Carroll2005, Easson2005}, the specific model in
which the correction is a polynomial of $R^{2},R^{ab}R_{ab}$ and
$R^{abcd}R_{abcd}$ is considered and the analysis there shows that
late-time accelerating attractor solutions exist. Meanwhile,
models with $R,R^{ab}R_{ab}$ corrections are discussed within the
Palatini approach \cite{Vollick2003, Allemandi2004a,
Allemandi2004b, Allemandi2005}, in which the field equations are
second order, and similar acceleration solutions are found. The
Palatini-$f(R)$ theory of gravitation was then tested using
various cosmological data such as Supernovae (SN), Cosmic
Microwave Background (CMB) shift parameter, baryon oscillation and
Big Bang Nucleosynthesis (BBN), in \cite{Capozziello2006,
Amarzguioui2006, Sotiriou2006, Fay2007, Borowiec2006}. Recently,
constraints from CMB and matter power spectra have also been given
in \cite{Koivisto2006, Li2006a, Li2006b}. These constraints
(especially that from the matter power spectrum) successfully
exclude most of the parameter space, making the model
indistinguishable from the standard $\Lambda \mathrm{CDM}$ in
practice.

Turning to the metric $f(R)$ gravity theories, the field equations
become fourth order and more difficult to handle \cite{BOtt}.
Until now, much attention has been focused on the solar system
tests of the theory and the existing results appear to exclude a
viable $f(R)$ cosmological model that leads to the current cosmic
accelerating expansion (see, for example, \cite{Chiba2003,
Erickcek2006, Navarro2006, Olmo2005a, Olmo2005b, Olmo2006} and
references therein; also see \cite{Zhang2007} and references
therein for arguments against these). Regardless of these local
considerations, however, we can look at the cosmological behavior
of the theory in order to arrive at an independent test. The
cosmological constraints on $f(R)$ gravity is relevant also
because one can simply imagine that the baryons do not see the
modifications to GR (a possibility suggested in
\cite{Amendola2006a}) evades the solar system tests in a somewhat
artificial way), or transform to the Einstein frame and consider a
scalar field model mathematically equivalent to the Jordan frame
$f(R)$ gravity \cite{Capozziello2006b}, but with the scalar field
coupling only to the dark matter. For further discussion of the
cosmology of the $f(R)$ dark energy models see, for example,
\cite{Odintsov2006a, Mena2006, Fay2007b} and references therein.

In a recent study, \cite{Amendola2006a}, the authors find that the class of
Friedmann cosmological models arising in a theory with $f(R)=R+\mu
^{2(n+1)}/(-R)^{n}$ and $n>0$ (note the difference from their original paper
because of our sign convention, $R<0$), which was thought to lead to the
late-time acceleration because the correction term to GR becomes significant
at late times, cannot reproduce the matter-dominated era of conventional
cosmology (yielding $a\propto t^{1/2}$ rather than $a\propto t^{2/3}$ where $%
a$ is the scale factor and $t$ the cosmic time) and so will be
ruled out by measurements of the redshift distance (see also
\cite{Fairbairn2007}). In this work they do not consider the case
of $-1<n<0$ which \emph{is} able to give a standard
matter-dominated era, as we show below. In a later work,
\cite{Amendola2006b}, these authors consider the cosmological
viabilities of general $f(R)$ models and have included this
possibility.

Since there are few known models with simple forms of $f(R)$
having the capability to consistently describing the whole
evolutionary history of the universe (see however \cite{Comment0}
for a discussion on this), we will also consider the perturbation
evolutions of the model. To this end we derive the covariant and
gauge invariant perturbation equations and put them into the
public CAMB code \cite {Lewis2000} in order to obtain the CMB and
matter power spectra. Note that the dynamics of perturbations for
$f(R)$ gravity have also been considered in \cite{Song2006,
Bean2006}. In \cite{Song2006} the background expansion is fixed to
match $\Lambda\mathrm{CDM}$ while in \cite{Bean2006} two specified
models are considered which do not give standard matter-dominated
eras and the calculation is done in the Einstein frame; thus both
analyses differ from the work reported here.

This paper is organised as the following: in
Section~\ref{sect:FEquations} we briefly introduce the
cosmological model and list the background and perturbation
evolution equations that are needed in the numerical calculation.
In Section~\ref{sect:Numerics} we solve for the background
evolution of the model numerically, and incorporate the
perturbations equations into CAMB in order to calculate the CMB
and matter power spectra. The discussion and conclusions are then
presented in Section~\ref{sect:Conclusion}. Throughout this work
we will set $c=$ $\hslash =1$ and only consider the case of a
spatially flat universe filled with photons, baryons, cold dark
matter (CDM) and three species of effectively massless neutrinos.

\section{Field Equations in f(R) Gravity}
\label{sect:FEquations}

In this section we very briefly summarise the main ingredients of
$f(R)$ gravity in metric approach, and list the general
perturbation equations that govern the dynamics of small
inhomogeneities in this theory.

\subsection{The Generalised Einstein Equations}

The starting point for the metric-$f(R)$ gravity is the
Einstein-Hilbert action,
\begin{eqnarray}
\label{eq:1}
S &=& \int d^{4}x\sqrt{-g}\left[ \frac{1}{2\kappa
}f(R)+\mathcal{L}_{m}\right] ,
\end{eqnarray}
in which $\kappa =8\pi G$ with $G$ being the gravitational
constant and $R = R(g_{ab})$ is the Ricci scalar. Varying the
action with respect to the metric $g_{ab}$ gives the modified
Einstein equations
\begin{eqnarray}
\label{eq:2}
FR_{ab}-\frac{1}{2}g_{ab}f(R)+(g_{ab}\nabla
^{c}\nabla _{c}-\nabla _{a}\nabla _{b})F &=& \kappa T_{ab},\ \
\end{eqnarray}
where $F=F(R)\equiv
\partial f(R)/\partial R$ and $T_{ab}$ is the energy-momentum
tensor. The trace of Eq.~(\ref{eq:2}) reads
\begin{eqnarray}
\label{eq:3}
FR-2f+3\nabla ^{c}\nabla _{c}F &=& \kappa T,
\end{eqnarray}%
with $T_{a}^{a}\equiv T=\rho -3p$. Let us call this the structural
equation, which relates $R$ (or $F$) directly to the energy
components in the universe. However, the term $\nabla ^{c}\nabla
_{c}F$ enters here which makes the equation a dynamical rather
than algebraic one. Recall, in contrast, that in the Palatini
$f(R)$ gravity the metric $g_{ab}$ and connection
$\Gamma^{a}_{bc}$ are treated as independent variables with $R =
R(g_{ab}, \Gamma^{a}_{bc})$; the variation of the action is taken
with respect to both of these two variables with the resulting
field equations being second order and the structural equation
simply an algebraic relation. Now we can also make a conformal
transformation,
\begin{eqnarray}
\label{eq:4}
\tilde{g}_{ab} &=& e^{2\omega }g_{ab},
\end{eqnarray}%
from the original Jordan frame to the Einstein frame (the tilded
quantities are the Einstein frame ones from here on), in which we
obtain the following action for the metric $f(R)$ gravity:
\begin{eqnarray}
\label{eq:5}
\tilde{S} &=& \int d^{4}x\sqrt{-\tilde{g}}\left[ -\frac{\tilde{R}(\tilde{g})}{%
2\kappa }-\frac{(\tilde{\nabla}\varphi )^{2}}{2}-V(\varphi )+\tilde{\mathcal{%
L}}_{m}\right] .\ \ \
\end{eqnarray}%
In Eq.~\ref{eq:5}, we have defined a new scalar field $\varphi \equiv \sqrt{6}%
\omega /\kappa \equiv \sqrt{3/2\kappa }\ln F$ and its potential
$V\equiv (f-FR)/2\kappa F^{2}$. Quantities in the Jordan and
Einstein frames are related to each other by \cite{Amendola2006a}
\begin{eqnarray}
\label{eq:6}
\tilde{\rho} = \rho e^{-4\omega },\tilde{p}=pe^{-4\omega },d\tilde{t}%
=e^{\omega }dt,\tilde{a}=e^{\omega }a. \end{eqnarray}%
We see that in the
Einstein frame the scalar field couples minimally to gravity but
couples conformally to matter, and that the generality in the
functional choice of $f(R)$ manifests itself in the generality of
the potential $V(\varphi )$.

Since we implement our measurements in the Jordan frame, we shall treat the
Jordan frame metric $g_{ab}$ as the physical one. Hence the difference
between $f(R)$ gravity and GR ($f(R)=R$) can be understood as a change in
the way that the spacetime curvature, and thus the physical Ricci tensor $%
R_{ab},$ responds to the distribution of matter.

\subsection{The Perturbation Equations}

The perturbation equations in general theories of $f(R)$ gravity derived in
this section which uses the method of $3+1$ decomposition \cite%
{Challinor1999, Lewis2000}. For more details, we refer the reader to these
references and only briefly outline its main ingredients before listing out
results.

The main idea of $3+1$ decomposition is to make spacetime splits
of physical quantities with respect to the 4-velocity $u^{a}$ of
an observer. The projection tensor $h_{ab}$ is defined as
$h_{ab}=g_{ab}-u_{a}u_{b}$ which can be used to obtain covariant
tensors perpendicular to $u$. For example, the covariant spatial
derivative $\hat{\nabla}$ of a tensor field $T_{d\cdot\cdot \cdot
e}^{b\cdot\cdot\cdot c}$ is defined as
\begin{eqnarray}
\label{eq:7} \hat{\nabla}^{a}T_{d\cdot\cdot\cdot
e}^{b\cdot\cdot\cdot c} &\equiv& h_{i}^{a}h_{j}^{b}\cdot\cdot\cdot
\ h_{k}^{c}h_{d}^{r}\cdot\cdot\cdot \ h_{e}^{s}\nabla
^{i}T_{r\cdot\cdot\cdot s}^{j\cdot\cdot\cdot k}.
\end{eqnarray}%
The energy-momentum tensor and covariant derivative of 4-velocity
are decomposed respectively as
\begin{eqnarray}
\label{eq:8}
T_{ab} &=&\pi _{ab}+2q_{(a}u_{b)}+\rho u_{a}u_{b}-ph_{ab}, \\
\label{eq:9}
\nabla _{a}u_{b} &=&\sigma _{ab}+\varpi
_{ab}+\frac{1}{3}\theta h_{ab}+u_{a}A_{b}.
\end{eqnarray}%
In the above, $\pi _{ab}$ is the projected symmetric trace-free (PSTF)
anisotropic stress, $q_{a}$ the vector heat flux vector, $p$ the isotropic
pressure, $\sigma _{ab}$ the PSTF shear tensor, $\varpi _{ab}=\hat{\nabla}%
_{[a}u_{b]}$, the vorticity, $\theta =\nabla ^{c}u_{c}=3\dot{a}/a$
($a$ is the mean scale factor) the expansion scalar, and
$A_{b}=\dot{u}_{b}$ the
acceleration; the overdot denotes time derivative expressed as $\dot{\phi}%
=u^{a}\nabla _{a}\phi $, brackets mean antisymmetrisation, and parentheses
symmetrization. The normalization is chosen as $u^{a}u_{a}=1$.

Decomposing the Riemann tensor and making use of the modified
Einstein equations, Eq.~(\ref{eq:2}) , we obtain, after
linearization, five constraint equations
\begin{eqnarray}
\label{eq:10}
0 &=&\hat{\nabla}^{c}(\varepsilon _{\ \ cd}^{ab}u^{d}\varpi _{ab}); \\
\label{eq:11}
\frac{\kappa q_{a}}{F} &=&\frac{\theta \hat{\nabla}_{a}F}{3F}-\frac{\hat{%
\nabla}_{a}\dot{F}}{F}-\frac{2\hat{\nabla}_{a}\theta }{3}+\hat{\nabla}%
^{b}\sigma _{ab}+\hat{\nabla}^{b}\varpi _{ab};\ \ \  \\
\label{eq:12}
\mathcal{B}_{ab} &=&\left[ \hat{\nabla}^{c}\sigma _{d(a}+\hat{\nabla}%
^{c}\varpi _{d(a}\right] \varepsilon _{b)ec}^{\ \ \ \ d}u^{e}; \\
\label{eq:13}
\hat{\nabla}^{b}\mathcal{E}_{ab} &=&\frac{1}{2F}\kappa \left[ \hat{\nabla}%
^{b}\pi _{ab}+\left( \frac{2}{3}\theta +\frac{\dot{F}}{F}\right) q_{a}+\frac{%
2}{3}\hat{\nabla}_{a}\rho \right]   \nonumber \\
&&-\frac{1}{2F^{2}}\left[ \kappa (\rho +p)\hat{\nabla}_{a}F-\dot{F}\hat{%
\nabla}_{a}\dot{F}+\ddot{F}\hat{\nabla}_{a}F\right] ; \\
\label{eq:14}
\hat{\nabla}^{b}\mathcal{B}_{ab} &=&\frac{1}{2F}\kappa \left[ \hat{\nabla}%
_{c}q_{d}+(\rho +p)\varpi _{cd}\right] \varepsilon _{ab}^{\ \
cd}u^{b}.
\end{eqnarray}%
Here, $\varepsilon _{abcd}$ is the covariant permutation tensor, $\mathcal{E}%
_{ab}$ and $\mathcal{B}_{ab}$ are respectively the electric and
magnetic parts of the Weyl tensor $\mathcal{W}_{abcd}$, given
respectively through $\mathcal{E}_{ab} =
u^{c}u^{d}\mathcal{W}_{acbd}$ and $\mathcal{B}_{ab} = -
\frac{1}{2}u^{c}u^{d}\varepsilon_{ac}^{\ \ ef}\mathcal{W}_{efbd}$.

In addition, we obtain seven propagation equations:
\begin{eqnarray}
\label{eq:15}
\dot{\rho}+(\rho +p)\theta +\hat{\nabla}^{a}q_{a} &=&0; \\
\label{eq:16}
\dot{q_{a}}+\frac{4}{3}\theta q_{a}+(\rho +p)A_{a}-\hat{\nabla}_{a}p+\hat{%
\nabla}^{b}\pi _{ab} &=&0; \\
\label{eq:17}
\dot{\theta}+\frac{1}{3}\left[ \theta -\frac{3\dot{F}}{F}\right] \theta -%
\hat{\nabla}^{a}A_{a}  \nonumber \\
-\left[ \frac{\hat{\nabla}^{2}F}{F}-\frac{\kappa \rho }{F}-\frac{f}{2F}%
\right]  &=& 0; \\
\label{eq:18}
\dot{\sigma}_{ab}+\frac{2}{3}\left[ \theta
+\frac{3\dot{F}}{4F}\right]
\sigma _{ab}-\hat{\nabla}_{\langle a}A_{b\rangle }  \nonumber \\
+\mathcal{E}_{ab}+\frac{1}{2F}\kappa \pi _{ab}+\frac{1}{2F}\hat{\nabla}%
_{\langle a}\hat{\nabla}_{b\rangle }F &=& 0; \\
\label{eq:19}
\dot{\varpi}+\frac{2}{3}\theta \varpi -\hat{\nabla}_{[a}A_{b]} &=&0; \\
\label{eq:20}
\frac{1}{2F}\kappa \left[ \dot{\pi}_{ab}+\left( \frac{1}{3}\theta -\frac{3%
\dot{F}}{2F}\right) \pi _{ab}\right]   \nonumber \\
-\frac{1}{2F}\kappa \left[ (\rho +p)\sigma _{ab}\
+\hat{\nabla}_{\langle
a}q_{b\rangle }\right]   \nonumber \\
-\frac{3\dot{F}}{4F^{2}}\hat{\nabla}_{\langle a}\hat{\nabla}_{b\rangle }F-%
\frac{3\dot{F}^{2}}{4F^{2}}\sigma _{ab}  \nonumber \\
-\left[ \dot{\mathcal{E}}_{ab}+\left( \theta
+\frac{\dot{F}}{2F}\right)
\mathcal{E}_{ab}-\hat{\nabla}^{c}\mathcal{B}_{d(a}\varepsilon
_{b)ec}^{\ \ \ \
d}u^{e}\right]  &=&0; \\
\label{eq:21}
\dot{\mathcal{B}}_{ab}+\left( \theta +\frac{\dot{F}}{2F}\right) \mathcal{B}%
_{ab}+\hat{\nabla}^{c}\mathcal{E}_{d(a}\varepsilon _{b)ec}^{\ \ \
\ d}u^{e} \nonumber \\
+\frac{1}{2F}\kappa \hat{\nabla}^{c}\mathcal{\pi
}_{d(a}\varepsilon _{b)ec}^{\ \ \ \ d}u^{e} &=&0,
\end{eqnarray}%
where the angle bracket means taking the trace-free part of a quantity.

Besides the above equations, it is useful to express the projected Ricci
scalar $\hat{R}$ into the hypersurfaces orthogonal to $u^{a}$ as
\begin{eqnarray}
\label{eq:22}
\hat{R} &\doteq& \frac{\kappa (\rho +3p)-f}{F}-\frac{2}{3}\theta ^{2}+\frac{1}{F%
}(\dot{F}\theta +3\ddot{F})+\frac{\hat{\nabla}^{2}F}{F}.
\end{eqnarray}%
The spatial derivative of the projected Ricci scalar, $\eta _{a}\equiv \frac{%
1}{2}a\hat{\nabla}_{a}\hat{R}$, is then given as
\begin{eqnarray}
\label{eq:23}
\eta _{a} &=& \frac{a}{F}\kappa \hat{\nabla}_{a}\rho +\frac{a}{F}\left[ \dot{%
\theta}+\frac{1}{3}\theta ^{2}\right]
\hat{\nabla}_{a}F-\frac{2a}{3}\left[
\theta +\frac{3\dot{F}}{2F}\right] \hat{\nabla}_{a}\theta   \nonumber \\
&&-\frac{a}{F}\theta \hat{\nabla}_{a}\dot{F}-\frac{a}{F}\hat{\nabla}_{a}\hat{%
\nabla}^{2}F,
\end{eqnarray}%
and its propagation equation by
\begin{eqnarray}
\label{eq:24}
\dot{\eta}_{a}+\frac{2\theta }{3}\eta _{a} &=& -\frac{a}{3F}\theta \hat{\nabla%
}_{a}\hat{\nabla}^{2}F-\left[ \frac{\dot{F}}{F}+\frac{2}{3}\theta \right] a%
\hat{\nabla}_{a}\hat{\nabla}\cdot A  \nonumber \\
&&-\frac{a}{F}\kappa \hat{\nabla}_{a}\hat{\nabla}\cdot q-\frac{a}{F}\hat{%
\nabla}_{a}(\hat{\nabla}^{2}F)^{\cdot }.
\end{eqnarray}

As we are considering a spatially flat universe, the spatial
curvature must vanish on large scales which means that
$\hat{R}=0$. Thus, from Eq.~(\ref{eq:22}), we obtain
\begin{eqnarray}
\label{eq:25}
\frac{1}{F}\kappa (\rho +3p)-\frac{2}{3}\theta ^{2}-\frac{f}{F}+\frac{\dot{F}%
}{F}\theta +\frac{3}{F}\ddot{F} &=& 0.
\end{eqnarray}%
This is just one of the modified Friedmann equations of the metric
$f(R)$ gravity, and the other modified background equations (the
other Friedmann equation and the energy-conservation equation) can
be obtained by taking the zero-order parts of Eqs.~(\ref{eq:17},
\ref{eq:15}), as
\begin{eqnarray}
\label{eq:26}
\dot{\theta}+\frac{1}{3}\left[ \theta -\frac{3\dot{F}}{F}\right] \theta +%
\frac{\kappa \rho }{F}+\frac{f}{2F} &=& 0; \\
\label{eq:27} \dot{\rho}+(\rho +p)\theta  &=& 0.
\end{eqnarray}%
It is easy to check that when $f(R)=R$, we have $F=1$ and these equations
reduce to those of GR.

To conclude this section, we want to point out another way to derive the
above perturbation equations. It follows by treating the new contributions
in the Einstein equation from the $f(R)$ gravity modification as an
additional effective energy momentum tensor \cite{Hwang1990}. In this way
one can write
\begin{eqnarray}
G_{ab} &=&\kappa \bar{T}_{ab}  \nonumber \\
&=&\frac{1}{F}\kappa T_{ab}+\frac{1}{F}\nabla _{a}\nabla _{b}F-\frac{1}{2F}%
(f+\kappa T-\nabla ^{2}F)g_{ab},  \nonumber
\end{eqnarray}%
where an overbar represents the total effective quantity, total effective
energy momentum tensor in this case. Now using the relations
\begin{eqnarray}
\rho  &=&T_{ab}u^{a}u^{b},  \nonumber \\
p &=&-\frac{1}{3}h^{ab}T_{ab},  \nonumber \\
q_{a} &=&h_{a}^{d}u^{c}T_{cd},  \nonumber \\
\pi _{ab} &=&h_{a}^{c}h_{b}^{d}T_{cd}+ph_{ab}  \nonumber
\end{eqnarray}%
we can identify the components of the total effective energy momentum tensor
as
\begin{eqnarray}
\kappa \bar{\rho} &=&\frac{\kappa }{2F}(\rho +3p)-\frac{f}{2F}+\frac{3\ddot{F%
}}{2F}+\frac{1}{2F}(\hat{\nabla}^{2}F+\dot{F}\theta ),  \nonumber \\
\kappa \bar{p} &=&\frac{\kappa }{2F}(\rho -p)+\frac{f}{2F}-\frac{5}{6F}(\hat{%
\nabla}^{2}F+\dot{F}\theta )-\frac{\ddot{F}}{2F},  \nonumber \\
\kappa \bar{q}_{a} &=&\frac{1}{F}\kappa q_{a}+\frac{1}{F}\hat{\nabla}_{a}%
\dot{F}-\frac{1}{3F}\theta \hat{\nabla}_{a}F,  \nonumber \\
\kappa \bar{\pi}_{ab} &=&\frac{1}{F}\kappa \pi _{ab}+\frac{1}{F}\hat{\nabla}%
_{\langle a}\hat{\nabla}_{b\rangle }F+\frac{\dot{F}}{F}\sigma _{ab}.
\nonumber
\end{eqnarray}%
The subsequent results follow exactly those well known in GR, just
with the components of the energy-momentum tensor being replaced
by the effective ones. We have checked this approach leads to the
same set of perturbation equations as we gave above.

\section{Numerical Results}
\label{sect:Numerics}

This section is devoted to the numerical calculations we have performed.
First, we obtain the background evolution of the present model, which we
discuss both in the Einstein frame and in the Jordan frame, and compare the
two. Once the background evolution at some time is known, we can implement
our set of perturbation equations on the CAMB code, with the background
quantities at arbitrary times (as required by the code) obtained by
interpolations. Some of the details are presented below.

\subsection{The Chameleon Mechanism in $f(R)$ Model}

The starting point for obtaining the background evolution is the
Eqs.~(\ref{eq:25} - \ref{eq:27}). However, before looking at this
calculation, we discuss the evolution of background quantities in
the Einstein frame, which provide us with a clearer physical
picture.

From the variation of the action, Eq.~(\ref{eq:5}), we can obtain
the following field equations:
\begin{eqnarray}
\label{eq:28}
\ddot{\varphi}+3\tilde{H}\dot{\varphi}+V_{\varphi } &=&\sqrt{\frac{\kappa }{6%
}}(\tilde{\rho}-3\tilde{p}), \\
\label{eq:29}
\dot{\tilde{\rho}}+3\tilde{H}(\tilde{\rho}+\tilde{p})
&=&-\sqrt{\frac{\kappa
}{6}}(\tilde{\rho}-3\tilde{p})\dot{\varphi},
\end{eqnarray}%
where $\tilde{H}=d\tilde{a}/\tilde{a}d\tilde{t}$, quantities $\tilde{\rho}$ and $\tilde{%
p}$ include contributions from both matter and radiation components, and $%
V_{\varphi }\equiv \partial V/\partial \varphi $. For the model
under consideration, $f(R)$ is given by
\[
f(R)=R+\frac{\mu ^{2(n+1)}}{(-R)^{n}},
\]%
so we have
\begin{eqnarray}
\label{eq:30}
F(R) &=& 1+n\left( \frac{\mu ^{2}}{-R}\right) ^{n+1}  \nonumber \\
&=& \exp \left( \sqrt{\frac{2\kappa }{3}}\varphi \right) ,
\end{eqnarray}%
and
\begin{eqnarray}
\label{eq:31}
V(\varphi ) &=& \frac{(n+1)\mu ^{2}}{2\kappa }\exp \left( -\frac{2}{3}\sqrt{%
6\kappa }\varphi \right)   \nonumber \\
&& \times \left[ \frac{\exp \left( \frac{\sqrt{6\kappa }}{3}\varphi \right) -1%
}{n}\right] ^{\frac{n}{n+1}}.
\end{eqnarray}%
Here, our definitions ensure that $[\exp (\sqrt{6\kappa }%
\varphi /3)-1]/n$ is always positive. At early times, when $\mu
^{2}/(-R)\ll 1,$ we have $F\rightarrow 1,\varphi \rightarrow 0$;
thus it follows immediately that when $-1<n<0$ we have, in this
limit, $F\rightarrow 1^{-}, \varphi \rightarrow 0^{-}, [\exp
(\sqrt{6\kappa }\varphi /3)-1]/n\rightarrow 0^{+}$ and so
$V\rightarrow \infty $. On the other hand, if $\varphi \rightarrow
-\infty$, we also have $V\rightarrow \infty $. Also, the potential
has a global minimum,
\[
V_{\mathrm{min}}(\varphi )=\frac{\mu ^{2}}{8\kappa (n+1)}(n+2)^{\frac{n+2}{%
n+1}},
\]%
at $\bar{\varphi}_{0}=\sqrt{\frac{3}{2\kappa }}\ln \frac{2n+2}{n+2}$.

Since there is a coupling to the matter, the evolution of $\varphi $ is not
determined solely by its potential, but by an effective potential $V_{%
\mathrm{eff}}(\varphi )$ given by (from Eq.~(\ref{eq:28}))
\[
V_{\mathrm{eff}}(\varphi )=V(\varphi )+\frac{1}{4}\rho _{\mathrm{m}}\exp
\left( -\frac{2\sqrt{6\kappa }}{3}\varphi \right) ,
\]%
in which $\rho _{\mathrm{m}}$ is the energy density of the
non-relativistic matter components, where we have used the
Jordan-frame matter density, which
is independent of $\varphi ,$ and the fact $p_{\mathrm{m}}=0, p_{\mathrm{rad}%
}=\rho _{\mathrm{rad}}/3$. The new minimum of the effective potential, $%
\varphi _{0}$, satisfies $\bar{\varphi}_{0}<\varphi _{0}<0$. From $\varphi
_{0}$ rightwards, the effective potential is dominated by $V(\varphi ),$
which is essentially a runaway one in this region; from $\varphi _{0}$
leftwards, it is dominated by the matter coupling, which increases
exponentially as $\varphi $ becomes more negative. Thus, this situation has
the advantageous properties of a Chameleon field \cite{Khoury2004a,
Khoury2004b, Mota2006a, Mota2006b}, the cosmology of which has been studied
in \cite{Brax2004a, Brax2004b}. However, there is one important difference
between our case and theirs. In order to see this, we show that the
effective mass-squared of small oscillations around the potential minimum, $%
m_{\varphi }^{2}=\partial ^{2}V_{\mathrm{eff}}(\varphi _{0})/\partial
\varphi ^{2}$, is given by
\[
m_{\varphi }^{2}\doteq -\frac{\kappa \rho _{\mathrm{m}}}{3n(n+1)}\left(
\frac{\kappa \rho _{\mathrm{m}}}{\mu ^{2}}\right) ^{n+1},
\]%
when $\sqrt{\kappa }|\varphi |$ is small enough (note that $n<0$).
Meanwhile, with $\sqrt{\kappa }|\varphi |\ll 1$ the quantities in the Jordan
and Einstein frames are essentially equal and we have
\begin{eqnarray}
\label{eq:32}
\frac{m_{\varphi }^{2}}{H^{2}} &\doteq& -\frac{\bar{\Omega}_{\mathrm{m}}}{n(n+1)%
}\left( \frac{\kappa \rho _{\mathrm{m}}}{\mu ^{2}}\right) ^{n+1},
\end{eqnarray}%
in which $\bar{\Omega}_{\mathrm{m}}\equiv \kappa \rho
_{\mathrm{m}}/3H^{2}$. During the matter-dominated era
$\bar{\Omega}_{\mathrm{m}}\sim 1$ and in the radiation dominated
era $\bar{\Omega}_{\mathrm{m}}\propto a$ while $(\kappa \rho
_{\mathrm{m}}/\mu ^{2})^{n+1}\propto a^{-3(n+1)}$. Thus, for
$n>-2/3,$ the ratio $m_{\varphi }^{2}/H^{2}$ always increases with
increasing redshift, and, because $(\kappa \rho _{\mathrm{m}}/\mu
^{2})^{n+1}\sim 1$ today, it is much larger than 1 at early times,
just as in the standard Chameleon scenario. The differences lies
in the fact that, at late times, in the present model $m_{\varphi
}^{2}/H^{2}\sim 1$ for $|n|$ not too close to $0$.

We can now depict the evolution history of the field as follows: at the
early times the effective mass of $\varphi $ is much heavier than $H$ so
that the field is attracted towards its effective potential minimum,
oscillates around it, but with the amplitude the oscillations being
gradually damped so that it eventually tracks the minimum. A similar
analysis to the one in ref. \cite{Brax2004a} can be made to show that the
field rolls slowly along this attractor. As the universe evolves, $%
m_{\varphi }/H$ gets smaller so that eventually the field begins to lag
behind its effective potential minimum and behaves like a normal
quintessence field, which contributes a dynamical dark energy. In the far
future, when $\rho _{\mathrm{m}}\rightarrow 0,$ the dynamics of the field is
determined by its potential only, so that it would evolve towards the
potential minimum, $\bar{\varphi}_{0},$ and stay there, after which the
universe commences de Sitter expansion in both the Einstein and Jordan
frames.

At this stage, we can also look at what happens if $n=0^{-}$.
Obviously, from Eq.~(\ref{eq:30}), $F=1^{-}$ and thus the field
$\varphi $ is confined to $0^{-}$ for all of the cosmic history.
From Eq.~(\ref{eq:32}), $m_{\varphi }^{2}/H^{2}\rightarrow \infty
$ so that any deviation from $0^{-}$ decays immediately. The
potential $V(\varphi )$ is, however, nonzero according to
Eq.~(\ref{eq:31}); in fact, in this limit we have
\[
\lim_{n\rightarrow 0^{-}}V=\frac{\mu ^{2}}{2\kappa }.
\]%
So, this represents a non-dynamical field with constant
potential, which is essentially a cosmological constant, and in
this case we recover the correct $\Lambda \mathrm{CDM}$ limit, as
expected.

Finally a comment on the solar system constraints on the $f(R)$
models. As discussed by the authors of Ref.~\cite{Faulkner2006},
when the baryons are allowed to see the $f(R)$ modification to GR,
the parameter region where one has a dynamical dark energy does
not overlap with that in which the chameleon effect could
successfully suppress the scalar-tensor type deviation from GR in
the solar system \cite{Comment2}. Here, as mentioned in
Sec.~\ref{sect:Intro}, however, we are merely concerned with
employing a "parametrized post-Friedman" description
\cite{Song2006} of the cosmological effects of the $f(R)$ model
and assume that these deviations from the standard gravitational
model do not affect the baryons. Furthermore, as noticed in
\cite{Navarro2006}, the exclusion of $f(R)$ models using solar
system constraints relies on the assumption that the transition
from GR-dominated dynamics to scalar-tensor dynamics on these
scales occurs adiabatically, which has not been investigated in
detail. The non-uniqueness of the static spherically symmetric
solutions to higher-order gravity theories is also a complicating
factor when determining the solar system bounds because of the
absence of a Birkhoff theorem; see Refs.~\cite{Clifton2005,
Clifton2006} for more details.

\subsection{Background Evolution}

Although we rely on the transformation to Einstein frame to obtain a
physical picture for the present model, our numerical calculations for the
background evolution will be implemented in the Jordan frame to be
consistent with later perturbation calculations. We describe these
calculations in this section.

Following \cite{Amendola2006b}, we make the following definitions
\begin{eqnarray}
\label{eq:33}
x_{1} &=&-\frac{\dot{F}}{FH}, \\
\label{eq:34}
x_{2} &=&\frac{f}{6FH^{2}}, \\
\label{eq:35}
x_{3} &=&-\frac{R}{6H^{2}}=\frac{\dot{H}}{H^{2}}+2, \\
\label{eq:36}
x_{4} &=&\Omega _{\mathrm{rad}}=\frac{\kappa \rho
_{\mathrm{rad}}}{3FH^{2}},\\
\label{eq:37}
m &=&\frac{RF_{R}}{F}, \\
\label{eq:38}
r &=&\frac{-RF}{f}=\frac{x_{3}}{x_{2}},
\end{eqnarray}%
where $F_{R}\equiv \partial F/\partial R=n(n+1)\mu
^{2(n+1)}/(-R)^{n+2}$, and, using the fact $R=-6(\dot{H}+2H^{2})$
at the background level, write Eqs.~(\ref{eq:25} - \ref{eq:27}) as
\begin{eqnarray}
\label{eq:39}
\frac{dx_{1}}{d\ln (a)} &=&-1-3x_{2}-x_{3}+x_{4}+x_{1}^{2}-x_{1}x_{3}; \\
\label{eq:40}
\frac{dx_{2}}{d\ln (a)} &=&x_{2}(x_{1}-2x_{3}+4)+\frac{x_{1}x_{3}}{m}; \\
\label{eq:41}
\frac{dx_{3}}{d\ln (a)} &=&2x_{3}(2-x_{3})-\frac{x_{1}x_{3}}{m}; \\
\label{eq:42}
\frac{dx_{4}}{d\ln (a)} &=&x_{4}(x_{1}-2x_{3}).
\end{eqnarray}

\begin{figure}[]
\includegraphics[scale=0.4]{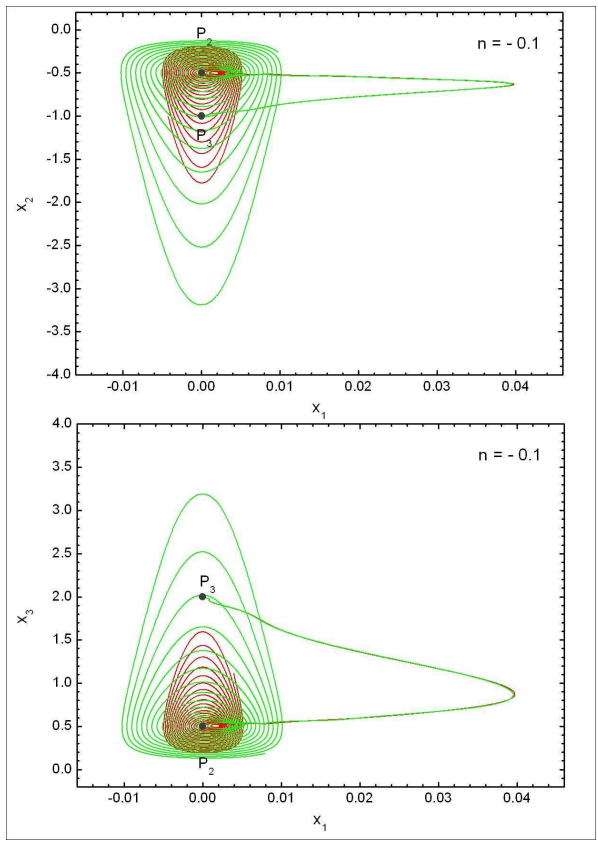}
\caption{(Color online) A phase portrait of the model. The
trajectories in the 3 dimensional space spanned by $x_{1}, x_{2},
x_{3}$ are projected to the $x_{1}-x_{2}$ and $x_{1}-x_{3}$ planes
respectively. The critical points $P_{2}, P_{3}$ are shown. Only 2
different trajectories with different initial conditions are
depicted for clearness and $n = - 0.1$ is adopted. It is seen that
$x_{1}$ oscillates around 0. Note that these are just for
illustrations and do not represent realistic cosmologies.}
\label{fig:1}
\end{figure}

The dynamics of this system has been thoroughly investigated in \cite%
{Amendola2006b}, here we just list the main results for the model $%
f(R)=R+\mu ^{2(n+1)}/(-R)^{n}$ for the purpose of completeness.
Firstly, it is easy to obtain
\begin{eqnarray}
\label{eq:43} \left(\frac{\mu ^{2}}{-R}\right) ^{n+1}&=&
\frac{x_{3}+x_{2}}{x_{3}-nx_{2}}
\end{eqnarray}%
for the use in the numerical calculation. Secondly, the critical
points and their properties of the system are listed in
Table~\ref{tab:table1} where we have defined
\begin{eqnarray}
\label{eq:44}
\Omega _{\mathrm{m}} &=&\frac{\kappa \rho _{\mathrm{m}}}{3FH^{2}}%
=1-x_{1}-x_{2}-x_{3}-x_{4}, \\
\label{eq:45}
w_{\mathrm{eff}} &=&-\frac{2\dot{H}}{3H^{2}}-1.
\end{eqnarray}%
\begin{table*}[tbp]
\caption{The critical points and their properties.}%
\begin{ruledtabular}
\begin{tabular}{ccccc}
Point & $(x_{1}, x_{2}, x_{3}, x_{4})$ & $\Omega_{_{m}}$ & $\Omega_{\mathrm{rad}}$ & $w_{\mathrm{eff}}$\\
\hline
$P_{1}$ & $(0,0,0,1)$ & 0 & 1 & $\frac{1}{3}$\\
$P_{2}$ & $(0,-\frac{1}{2},\frac{1}{2},0)$ & 1 & 0 & 0\\
$P_{3}$ & $(0,-1,2,0)$ & 0 & 0 & $-1$\\
$P_{4}$ & $(-1,0,0,0)$ & 2 & 0 & $\frac{1}{3}$\\
$P_{5}$ & $(1,0,0,0)$ & 0 & 0 & $\frac{1}{3}$\\
$P_{6}$ & $\left(\frac{4(n+1)}{n},\frac{2(n+1)}{n^{2}},\frac{2(n+1)}{n},-\frac{5n^2+8n+2}{n^2}\right)$ & 0 & $-\frac{5n^2+8n+2}{n^2}$ & $-\frac{3n+4}{3n}$\\
$P_{7}$ & $\left(\frac{3(n+1)}{n},\frac{4n+3}{2n^{2}},\frac{4n+3}{2n},0\right)$ & $-\frac{8n^{2}+13n+3}{2n^{2}}$ & 0 & $-\frac{n+1}{n}$\\
$P_{8}$ & $\left(-\frac{2(n+2)}{2n+1},\frac{4n+5}{(2n+1)(n+1)},\frac{n(4n+5)}{(2n+1)(n+1)},0\right)$ & 0 & 0 & $-\frac{6n^{2}+7n-1}{3(2n+1)(n+1)}$\\
$P_{9}$ & $(-4,5,0,0)$ & 0 & 0 & $\frac{1}{3}$\\
\end{tabular}
\end{ruledtabular}
\label{tab:table1}
\end{table*}

It is easy to see from Table~\ref{tab:table1} that the points
$P_{1},P_{2},P_{3}$ correspond to a radiation-dominated era, a
matter-dominated era and a de Sitter era, respectively, while
$P_{4},P_{5},P_{9}$ cannot lead to any of these three eras and are
excluded from further consideration. In Table~\ref{tab:table2}, we
summarise the stability and acceleration properties of the
critical points other than $P_{4},P_{5},P_{9}$ (limited to the
cases in which $-1<n<0$). From this table we see that, for general
$-1<n<0,$ we can expect $P_{1},P_{2},P_{3}$ to give a viable
cosmology provided that the initial conditions are chosen
appropriately. In particular, around $P_{2}$, the evolutionary
trajectory of the model is a damped oscillation, and it is finally
attracted to the stable de Sitter point $P_{3}$. In
Figure~\ref{fig:1} we have plotted a phase portrait to illustrate
this evolution: we see that evolutions with rather different
initial conditions fall towards the vicinity of $P_{2}$ and
finally pass it, being attracted to the de Sitter point $P_{3}$
along nearly the same trajectory (with this trajectory itself
depending on the value of $n$). Depending on the initial
conditions, the region of the oscillation could be very tiny; but
the numerical simulation shows that, the oscillation of $x_{1}$
around $0$ is a general result, at least for some period of the
evolution.

\begin{figure*}[]
\includegraphics[scale=0.85]{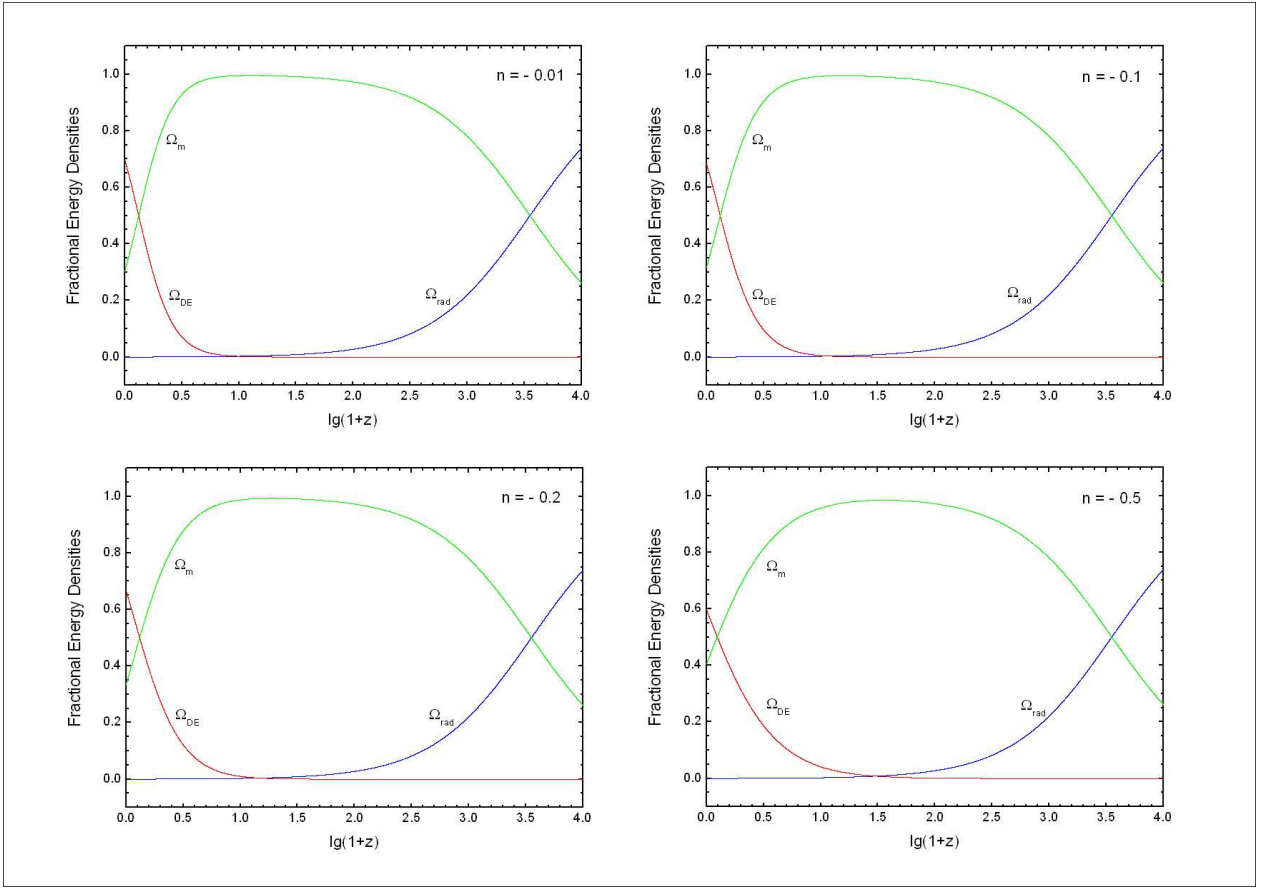}
\caption{(Color online) The fractional energy densities
$\Omega_{\mathrm{m}}$ (green lines), $\Omega_{\mathrm{rad}}$ (blue
lines) and $\Omega_{\mathrm{DE}}\equiv
1-\Omega_{\mathrm{m}}-\Omega_{\mathrm{rad}}$ (red lines) as
functions of redshift $z$. We plot the cases for $n = -0.01, -0.1,
-0.2$ and $-0.5$ respectively. Note that in general
$\Omega_{\mathrm{m0}} > 0.3$ because we require
$\kappa\rho_{\mathrm{m0}}/3H_{0}^{2} = 0.3$ instead and because
$F_{0} < 1$.} \label{fig:2}
\end{figure*}

\begin{figure}[]
\includegraphics[scale=0.5]{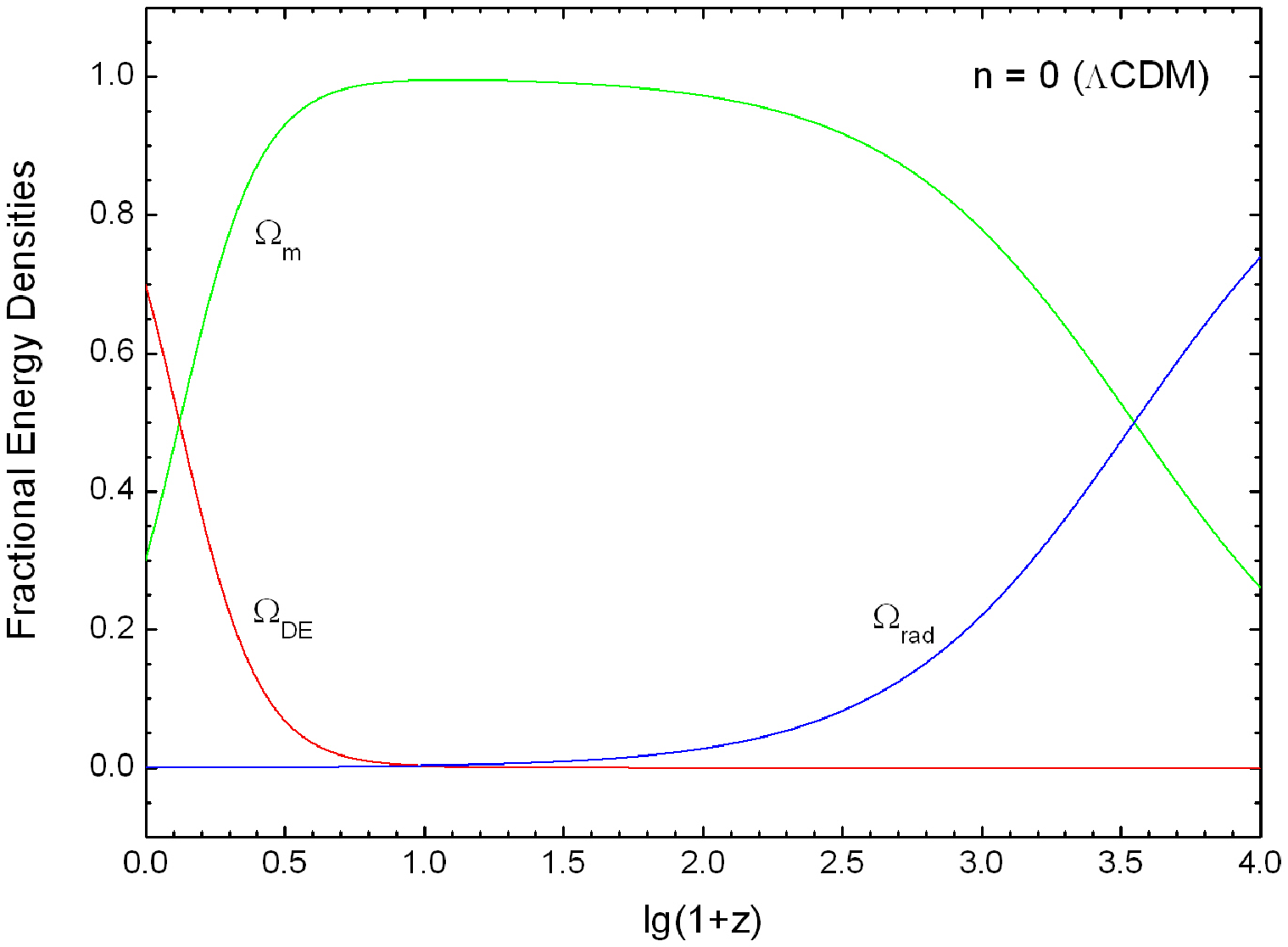}
\caption{(Color online) The same as FIG.~2, but for the case of
$n=0$ ($\Lambda\mathrm{CDM}$).} \label{fig:3}
\end{figure}

One can recognise here the connection with our Einstein frame
analysis. Note that $\varphi \propto \ln F$, so
$\dot{\varphi}\propto \dot{F}/F$ and an oscillation of $x_{1}$
around $0$ corresponds to an oscillation of $\dot{F}$ around $0$,
which in turn represents an oscillation of $\dot{\varphi}$ around
$0$ (this is just the condition for a $\varphi $ oscillation).
Furthermore, the final attraction of the trajectories towards
$P_{3}$ is also consistent with our conclusion above, that the
universe will finally evolve into a de Sitter stage.
\begin{table*}[tbp]
\caption{The stability and acceleration properties of the critical points ($%
-1<n<0$).}%
\begin{ruledtabular}
\begin{tabular}{cccc}
Point & Eigenvalues & Stability & Acceleration ?\\
\hline
$P_{1}$ & $1,4,4,-1$ & Saddle & No \\
$P_{2}$ & $-1, \left[3+\frac{3n}{(1+m)^{2}}, -\frac{3}{4}\pm\sqrt{-\frac{1}{m}}\right]_{m\rightarrow 0^{+}}$ & Saddle & No \\
$P_{3}$ & $-3,-4,-\frac{3}{2}\pm\frac{\sqrt{25+\frac{32}{n}}}{2}$ & Stable & Yes\\
$P_{6}$ & $1,4\left(1+\frac{1}{n}\right),\frac{1}{2}+\frac{1}{n}\pm\frac{\sqrt{81n^{2}+132n+36}}{2n}$ & Saddle & No\\
$P_{7}$ & $-1,3\left(1+\frac{1}{n}\right),\frac{3(n+1)\pm\sqrt{256n^4+864n^3+1025n^2+498n+81}}{4n(n+1)}$
& Stable if $-\frac{3}{4}\leq n \leq \frac{\sqrt{73}-13}{16}$, saddle otherwise \footnotemark[1] & No \\
$P_{8}$ & $-\frac{4n+5}{n+1},-\frac{2(n+2)}{2n+1},-\frac{2(5n^2+8n+2)}{(2n+1)(n+1)},-\frac{8n^2+13n+3}{(2n+1)(n+1)}$
& Stable if $n \geq \frac{\sqrt{73}-13}{16}$, saddle otherwise & $-1<n<-\frac{1}{2}$ \footnotemark[2] \\
\end{tabular}
\end{ruledtabular}
\footnotetext[1]{As $n\rightarrow -1$ this gives a matter era, but
one of the eigenvalues becomes $\infty$ so that the system cannot
stay around the point for a long time. Also in this case $f(R) = R
+ \alpha/(-R)^{n} \propto R$, meaning that the modified $f(R)$
gravity is merely GR with a different gravitational
constant.}\\
\footnotetext[2]{Strong phantom era with $w_{\mathrm{eff}}<-7.6$;
unstable.} \label{tab:table2}
\end{table*}

\begin{figure}[]
\includegraphics[scale=0.5]{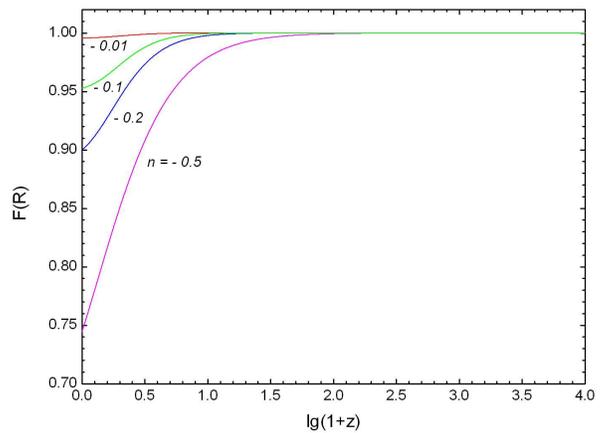}
\caption{(Color online) $F$ as a function of redshift. The red,
green, blue and magenta curves are the cases for $n = -0.01, -0.1,
-0.2$ and $-0.5$ respectively).} \label{fig:4}
\end{figure}

In order to go further, and solve the background equations
numerically, we need the detailed initial conditions. One possible
way to obtain these is to consider the evolution in the
radiation-dominated era. For example, it is
possible (in Einstein frame) that the field is frozen in this era \cite%
{Brax2004a} while settling in the vicinity of its effective potential
minimum and then starting to oscillate as soon as the fractional matter
energy density becomes significant. In our calculation, we use a trial and
error method to find the initial conditions at some specified early time
which reproduce the observational value $\kappa \rho _{\mathrm{m0}%
}/3H_{0}^{2}\simeq 0.3$ and $\kappa \rho
_{\mathrm{rad0}}/3H_{0}^{2}\simeq 10^{-4},$ where subscript '0'
denotes the present-day value (note that we
are not using $\Omega _{\mathrm{m0}}\simeq 0.3,\Omega _{\mathrm{rad0}%
}\simeq 10^{-4}$ because of the different definitions
Eqs.~(\ref{eq:36}, \ref{eq:44}) from conventional cosmology). The
results for some specific choices of $n$ are given in
Figure~\ref{fig:2}. Although the difference is not large in the
plots, we can see that the $f(R)$ dark energy starts to be
significant earlier for larger $|n|$'s. This is as expected
because the ratio of correction term in $f(R)$ to $-R$ is given by
$[\mu ^{2}/(-R)]^{n+1}$ which becomes of order $1$ earlier if
$n+1$ is closer to $0$. On the other hand, evolutions of the
fractional energy densities are essentially the same as that for
$\Lambda \mathrm{CDM}$ at early times when the $f(R)$ corrections
are negligible. For reference, we also show the case of $\Lambda
\mathrm{CDM}\ (n=0)$ in Figure~\ref{fig:3}.

Next, we outline the way to obtain all other background quantities
from what has already been calculated. Firstly, from
Eq.~(\ref{eq:43}) we now have
\begin{eqnarray}
\label{eq:46}
F &=& (1+n)\frac{x_{3}}{x_{3}-nx_{2}}.
\end{eqnarray}%
Secondly, from Eq.~(\ref{eq:36}) we get
\begin{eqnarray}
\frac{x_{4}}{x_{40}} &=&
\frac{F_{0}H_{0}^{2}}{FH^{2}}a^{-4},\nonumber
\end{eqnarray}%
so that
\begin{eqnarray}
\label{eq:47} H^{2} &=& \frac{F_{0}x_{40}}{Fx_{4}}a^{-4}H_{0}^{2}.
\end{eqnarray}%
From Eqs.~(\ref{eq:33}, \ref{eq:35}) it follows that
\begin{eqnarray}
\label{eq:48}
\dot{F} &=&-x_{1}FH, \\
\label{eq:49}
R &=&-6x_{3}H^{2}, \\
\label{eq:50} \dot{H} &=&(x_{3}-2)H^{2}.
\end{eqnarray}%
Finally, Eqs.~(\ref{eq:25}, \ref{eq:26}) give
\begin{eqnarray}
\label{eq:51} \ddot{F} &=& (1+2x_{1}+3x_{2}+x_{3}-x_{4})FH^{2}.
\end{eqnarray}%
Consequently, provided $H_{0}$ is known, we can compute any other background
quantity.

The value of $F$ plays an important role in the $f(R)$ cosmology and so we
discuss it separately here. During the matter-dominated era, the trajectory
oscillates around the point $P_{2}$, at which $F=1$ exactly. The relation $%
x_{2}+x_{3}=0$ holds approximately so that $F$ remains close to $1$. At the
de Sitter attractor $P_{3}$, we have $x_{2}=-1$ and $x_{3}=2,$ so that $%
F=(2n+2)/(n+2)$ (corresponding to the potential minimum at
$\bar{\varphi}_{0} $ in the Einstein frame). Since our universe
has not yet reached $P_{3}$, we expect that $(2n+2)/(n+2)<F_{0}<1$
at present, which is confirmed in Figure~\ref{fig:4}, where we
plot the time evolution of $F$ for the same choices of $n$ as
above.

Thus far we have calculated the background quantities $x_{1},
x_{2}, x_{3}, x_{4} $ at some predefined time grid points (scale
factor values). Then, in the CAMB code, we can obtain the
background quantities at arbitrary times (scale factor values)
simply by interpolating between these grid points.

\subsection{The CMB and Matter Power Spectra}

In the metric $f(R)$ gravity theory, $F$ and $R$ are determined by the
energy-momentum tensor $T_{ab}$ through a dynamic equation, as are the
perturbations of them, $\hat{\nabla}_{a}F$ or $\hat{\nabla}_{a}R$. To see
this, take the covariant spatial derivative of the structural equation and
we get
\begin{eqnarray}
\label{eq:52}
\kappa (\hat{\nabla}_{a}\rho -3\hat{\nabla}_{a}p) &=&3(\hat{\nabla}_{a}\ddot{%
F}+\theta \hat{\nabla}_{a}\dot{F}+\dot{F}\hat{\nabla}_{a}\theta +\hat{\nabla}%
_{a}\hat{\nabla}^{2}F)  \nonumber \\
&&+R\hat{\nabla}_{a}F-\frac{F}{F_{R}}\hat{\nabla}_{a}F,
\end{eqnarray}%
which should be added to our set of perturbation equations. Since this is a
second-order differential equation, we can recast it into two first-order
differential equations, which means that we have two more quantities to
evolve in the CAMB code, whose initial conditions also need to be specified.

\begin{figure}[]
\includegraphics[scale=0.5]{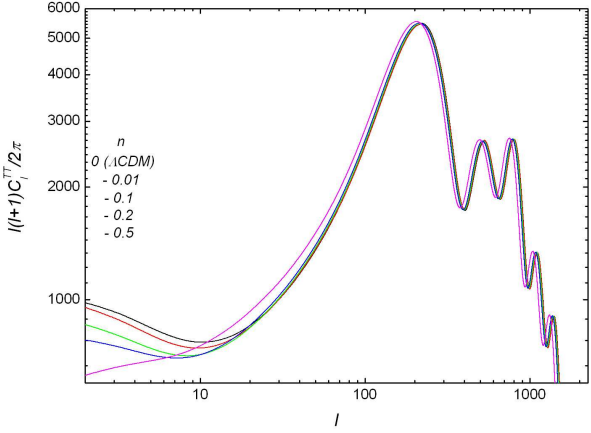}
\caption{(Color online) The CMB power spectra of the $f(R)$ model.
The black, red, green, blue and magenta lines represent the cases
of $n = 0, -0.01, -0.1, -0.2$ and $-0.5$ respectively. For all the
curves we have adopted the parameters $H_{0} = 70\
\mathrm{km/s/Mpc}$ and $\kappa\rho_{\mathrm{m0}}/3H_{0}^{2} =
0.3$.} \label{fig:5}
\end{figure}

\begin{figure}[]
\includegraphics[scale=0.5]{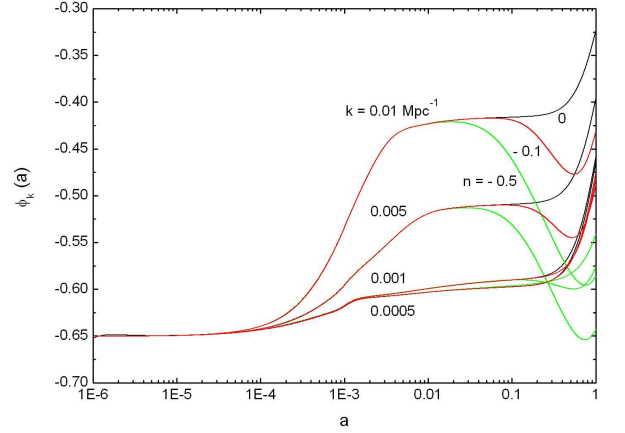}
\caption{(Color online) The potentials $\phi_{k}$ (see definition
in text) as functions of the scale factor $a$ for various $k$
values. Only the cases of $\Lambda\mathrm{CDM}\ (n=0)$ (black
curves), $n = -0.2$ (red curves) and $n = -0.5$ (green curves) are
shown for clearness. The scales are chosen to be $k = 0.0005,
0.001, 0.005, 0.01\ \mathrm{Mpc}^{-1}$ as indicated beside the
curves. For all the curves we have adopted the parameters $H_{0} =
70\ \mathrm{km/s/Mpc}$ and $\kappa\rho_{\mathrm{m0}} = 0.3$.}
\label{fig:6}
\end{figure}

Making a harmonic expansion of $\hat{\nabla}_{a}F$ as
\[
\hat{\nabla}_{a}F=\sum_{k}\frac{k\epsilon }{a}Q_{a}^{k},
\]%
where $Q_{a}^{k}=\frac{a}{k}\hat{\nabla}_{a}Q^{k}$ and $Q^{k}$ are the
zero-order eigenvalues of the comoving Laplacian $a^{2}\hat{\nabla}^{2}$ so
that $a^{2}\hat{\nabla}^{2}Q^{k}=k^{2}Q^{k}$ \cite{Lewis2000}, it is easy to
show that
\begin{eqnarray}
\label{eq:53}
\hat{\nabla}_{a}\dot{F} &=&\sum_{k}\frac{k\dot{\epsilon}}{a}Q_{a}^{k}, \\
\label{eq:54} \hat{\nabla}_{a}\ddot{F}
&=&\sum_{k}\frac{k\ddot{\epsilon}}{a}Q_{a}^{k},
\end{eqnarray}%
with the aid of
\[
\hat{\nabla}_{a}\dot{\psi}\doteq (\hat{\nabla}_{a}\psi )^{\cdot }+\frac{1}{3}%
\theta \hat{\nabla}_{a}\psi
\]%
for any scalar $\psi $ (the $\doteq $ means that we obtain this
relation in the frame where the acceleration $A_{a}=0$, see
\cite{Lewis2000}). At early times, when $F=1,\dot{F}=0$,
Eq.~(\ref{eq:52}) reduces to
\begin{eqnarray}
\label{eq:55}
&&\frac{\kappa (\mathcal{X}-3\mathcal{X}^{p})}{3H^{2}}  \nonumber \\
&\approx & \epsilon ^{\star \star }+\frac{3}{2}\epsilon ^{\star }
+ \left[\frac{R}{3H^{2}}\left( 1-\frac{1}{RF_{R}}\right) +\frac{k^{2}}{a^{2}H^{2}}%
\right] \epsilon ,
\end{eqnarray}%
in which $\mathcal{X},\mathcal{X}^{p}$ are defined through the harmonic
expansions $\hat{\nabla}_{a}\rho =\sum_{k}k\mathcal{X}Q_{a}^{k}/a$ and $\hat{%
\nabla}_{a}p=\sum_{k}k\mathcal{X}^{p}Q_{a}^{k}/a$, and a star denotes the
derivative with respect to $\ln a$. Recall that in the matter-dominated era $%
|R/H^{2}|\sim 1$, $\kappa
(\mathcal{X}-3\mathcal{X}^{p})/3H^{2}=\kappa \rho
_{\mathrm{m}}\Delta _{\mathrm{m}}/3H^{2}\sim \Delta
_{\mathrm{m}}$, where $\Delta _{\mathcal{\mathrm{m}}}$ is the
fractional perturbation in non-relativistic matter components (the
contribution from photons and massless neutrinos to this term is
zero because $\mathcal{X}_{\gamma ,\nu }^{p}=\mathcal{X}_{\gamma
,\nu }/3$) and
\[
1-\frac{1}{RF_{R}}=1+\frac{1}{n(n+1)\left( \frac{\mu ^{2}}{-R}\right) ^{n+1}}
\]%
tends to $-\infty $ (for $-1<n<0$) as $\mu ^{2}/(-R)\rightarrow 0$. So,
early in the matter-dominated era, the coefficient in front of $\epsilon $
becomes very large (remember again that $R<0$) and $\epsilon $ settles to $%
\sim \Delta _{\mathrm{m}}/|1-1/RF_{R}|$ within a short time \cite{Song2006}.
This means that the actual result will be insensitive to the choice of
initial conditions, a fact we have checked using the numerical code. Thus,
for our calculation, we can simply set $\epsilon _{\mathrm{initial}}=\dot{%
\epsilon}_{\mathrm{initial}}=0$. In this model, the perturbation in $F$ is
driven essentially to zero (compared with the matter perturbation) and
finally grows as dark energy starts to drive the expansion of the universe.
It is also interesting to note that according to the analysis above, when $%
n=0^{-}$ we have $F_{R}=0^{-}$ and so the perturbation,
$\epsilon$, will stay zero all through cosmic history, which is
consistent with the property of an effective cosmological
constant.

In Figure~\ref{fig:5}, we plot the CMB power spectrum of the model
for different
choices of $n$. For all of these plots we again adopt $H_{0}=70\ \mathrm{%
km/s/Mpc}$ and $\kappa \rho _{\mathrm{m0}}/3H_{0}^{2}=0.3$. Two effects can
immediately be seen to occur in the spectrum. Firstly, the locations of the
acoustic peaks move towards the lower $l$s as $|n|$ increases. This is
because, for larger $|n|,$ the dark-energy term starts to be important
earlier and, with $H_{0}$ fixed, the universe  has a smaller age today,
which causes the peaks to shift leftwards. This effect is not very
significant when $n$ is small. Secondly, we see a reduction of power at low $%
l$s, which was also been found and discussed in \cite{Song2006},
although there the background expansion is fixed to match the
$\Lambda \mathrm{CDM}$ cosmology. This reduction in low-$l$ power
is caused by a weaker late-time decay of the large-scale potential
$\phi _{k}$ (which is the coefficient of the harmonic expansion of $%
\mathcal{E}_{ab}$ as $\mathcal{E}_{ab} = - \sum_{k}k^{2}\phi
_{k}Q_{ab}^{k}/a^{2}$ and related to the Newtonian potential
$\Psi$ by $\Psi = \phi - \kappa\Pi a^{2}/2k^{2}$ for any specified
$k$-mode where $\Pi$ is the anisotropic stress \cite{Lewis2000})
compared with that in $\Lambda \mathrm{CDM}$, that leads to a
suppression on the Integrated Sachs-Wolfe (ISW) effect and thus of
the $C_{l}$ at low $l$s. In Figure~\ref{fig:6} we give some plots
for the evolution of the potential at different scales ($k$'s) in
this model, which clearly show the slower decay at large scales
\cite{Comment1}. Note that this is remarkably different from the
cases of Palatini $f(R)$ gravities, in
which the potentials decay much more rapidly than that in $\Lambda \mathrm{%
CDM}$ \cite{Li2006a, Li2006b}, leading to significant amplifications of the
low-$l$ power. Also, we see that this model is potentially useful in
reducing the difference between the predicted and measured CMB\ powers at
low $l$.

\begin{figure}[]
\includegraphics[scale=0.5]{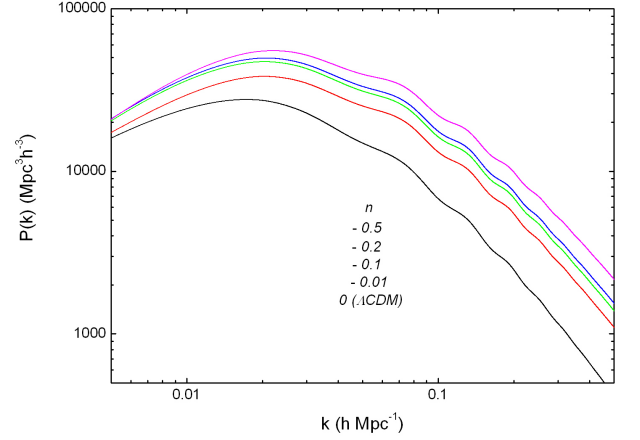}
\caption{(Color online) The matter power spectra of the $f(R)$
model. The black, red, green, blue and magenta lines represent the
cases of $n = 0, -0.01, -0.1, -0.2$ and $-0.5$ respectively. In
all these plots we have adopted $H_{0} = 70\ \mathrm{km/s/Mpc}$
and $\kappa\rho_{\mathrm{m0}}/3H_{0}^{2} = 0.3$.} \label{fig:7}
\end{figure}

The linear matter power spectra at $z=0$ of the present models are
shown in Figure~\ref{fig:7}. From this plot we can see that the
matter power spectrum has
similar behavior to those in the case of a general $f(R)$ model with $%
\Lambda \mathrm{CDM}$ background evolution \cite{Song2006}. On small scales
the spectra arising in the $f(R)$ gravity theories we are considering have
similar shapes to the case of the $\Lambda \mathrm{CDM}$ power spectrum, and
this fact can be understood roughly, as follows. Consider for simplicity the
growth of the dark-matter fractional density perturbation $\Delta _{\mathrm{c%
}},$ which is defined by $\rho _{\mathrm{c}}\Delta _{\mathrm{c}}=\mathcal{X}%
_{c}$ (essentially the $\delta _{\mathrm{c}}$ in CMBFAST), in a universe
filled with cold dark matter and photons. After some manipulations of our
set of perturbation equations it is easy to show that in metric $f(R)$
theories:
\begin{eqnarray}
\label{eq:56}
&&\Delta _{\mathrm{c}}^{\prime \prime }+\left( \mathcal{H}-\frac{F^{\prime }%
}{F}\right) \Delta _{\mathrm{c}}^{\prime }  \nonumber \\
&=&\frac{1}{F}\kappa \mathcal{X}a^{2}-\frac{1}{F}k^{2}\epsilon -\frac{3}{F}%
\mathcal{H}\epsilon ^{\prime }-\frac{3}{2F}\frac{\mathcal{H}^{\prime \prime
}+\frac{F^{\prime }}{F}\mathcal{H}^{\prime }}{\mathcal{H}+\frac{F^{\prime }}{%
2F}}\epsilon   \nonumber \\
&&+\frac{\frac{3\mathcal{H}}{F}(\mathcal{H}^{\prime }+\mathcal{H}^{2})\left(
1-\frac{F}{RF_{R}}\right) }{\mathcal{H}+\frac{F^{\prime }}{2F}}\epsilon
\end{eqnarray}%
where a prime represents the derivative with the conformal time $\tau $ ($%
ad\tau =dt$) and $\mathcal{H}=a^{\prime }/a$. At the same time,
Eq.~(\ref{eq:52}) can now be rewritten as
\begin{eqnarray}
\label{eq:57}
\kappa (\mathcal{X}-3\mathcal{X}^{p})a^{2} &=& 3(\epsilon ^{\prime \prime }+2%
\mathcal{H}\epsilon ^{\prime }+kF^{\prime }\mathcal{Z}+k^{2}\epsilon )
\nonumber \\
&& -6(\mathcal{H}^{\prime }+\mathcal{H}^{2})\left(
1-\frac{F}{RF_{R}}\right) \epsilon
\end{eqnarray}%
where $\mathcal{Z}$ is the harmonic expansion coefficient of $\hat{\nabla}%
_{a}\theta $ ($\hat{\nabla}_{a}\theta =\sum_{k}k^{2}\mathcal{Z}%
Q_{a}^{k}/a^{2}$). In the CDM frame (in which $v_{c}=0$) we have \cite%
{Lewis2000}
\begin{eqnarray}
\label{eq:58} \Delta _{\mathrm{c}}^{\prime } &=& -k\mathcal{Z}.
\end{eqnarray}%
A similar analysis to that below Eq.~(\ref{eq:55}) then shows
that, for small scales
which are characterised by $k^{2}/\mathcal{H}^{2}\gg 1$, the term $%
3k^{2}\epsilon $ dominates the right-hand side of Eq.~(\ref{eq:57}) provided that $%
|F_{R}|$ is not too close to $0$, which is the case for the later stages in
our $f(R)$ cosmology. Thus, we have now
\begin{eqnarray}
\label{eq:59}
\kappa (\mathcal{X}-3\mathcal{X}^{p})a^{2} &\approx& 3(kF^{\prime }\mathcal{Z}%
+k^{2}\epsilon ),
\end{eqnarray}%
which we have also verified numerically.

Substituting Eqs.~(\ref{eq:58}, \ref{eq:59}) into
Eq.~(\ref{eq:56}), we get (neglecting the contribution from
relativistic energy components):
\begin{eqnarray}
\label{eq:60}
\Delta _{\mathrm{c}}^{\prime \prime }+\mathcal{H}\Delta _{\mathrm{c}%
}^{\prime }-\frac{2}{3F}\kappa \rho _{\mathrm{c}}\Delta _{\mathrm{c}%
}a^{2}&\approx& 0.
\end{eqnarray}%
Interestingly, we have obtained a scale-independent behavior for the CDM
density perturbation growth on small scales, similar to that of standard GR,
where it is given by
\begin{eqnarray}
\label{eq:61}
\Delta _{\mathrm{c}}^{\prime \prime }+\mathcal{H}\Delta _{\mathrm{c}%
}^{\prime }-\frac{1}{2}\kappa \rho _{\mathrm{c}}\Delta _{\mathrm{c}%
}a^{2}&\approx& 0.
\end{eqnarray}%
This explains why on small scales the shape of the $f(R)$ matter power
spectrum is like that of $\Lambda \mathrm{CDM}$ (see \cite{Koivisto2006b}
for a similar example). It also indicates that the growths of small-scale
perturbations are governed by an effective gravitational constant given by $%
\kappa _{\mathrm{eff}}=4\kappa /3F>\kappa $. Notice that one
should not simply try to put $F=1$ into Eq.~(\ref{eq:60}) to get
the $\Lambda \mathrm{CDM}$ limit, since under this condition the
last term in Eq.~(\ref{eq:57}) always dominates over
$3k^{2}\epsilon $ and can never be neglected. The correct $\Lambda
\mathrm{CDM}$ limit is recovered by setting $F^{\prime }=0,F=1,6(\mathcal{H}%
^{\prime }+\mathcal{H}^{2})(1-F/RF_{R})\epsilon =-\kappa (\mathcal{X}-3%
\mathcal{X}^{p})a^{2},$ (according to Eq.~(\ref{eq:57})) and all
other terms relating $\epsilon $ and $\epsilon ^{\prime }$ to $0$
in Eq.~(\ref{eq:56}), which just leads to Eq.~(\ref{eq:61}), and
shows that the early-matter-era growth of perturbations is well
described by GR.

\begin{figure}[]
\includegraphics[scale=0.45]{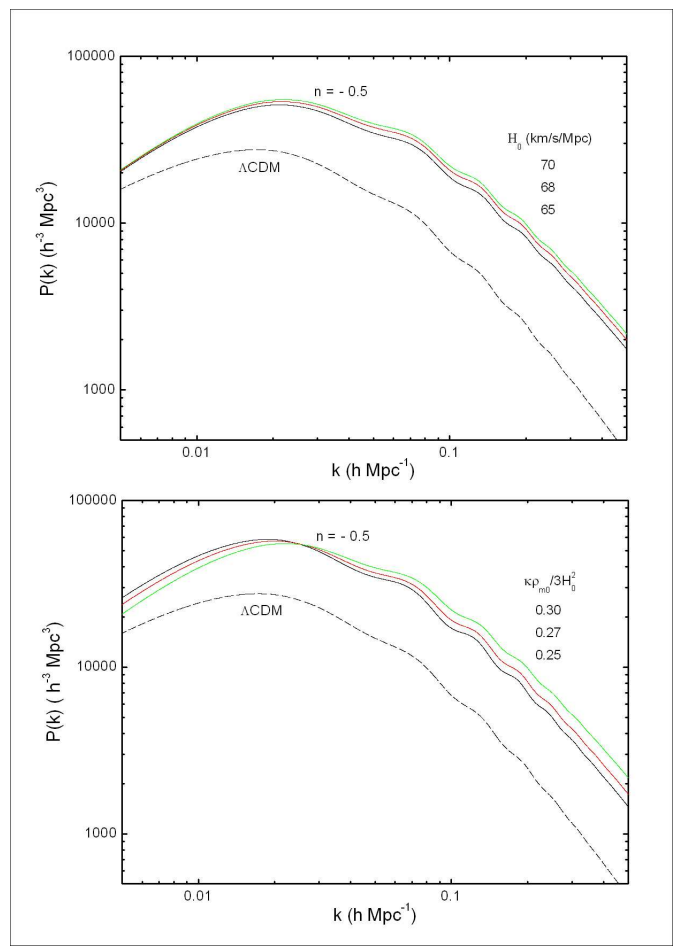}
\caption{(Color online) The dependence of the matter power spectra
on $H_{0}$ (upper panel) and $\kappa\rho_{\mathrm{m0}}/3H_{0}^{2}$
(lower panel). Upper panel: the black, red, green solid curves
represent $H_{0} = 65, 68, 70\ \mathrm{km/s/Mpc}$ respectively
($\kappa\rho_{\mathrm{m0}}/3H_{0}^{2} = 0.3$). Lower panel: the
black, red, green curves represent
$\kappa\rho_{\mathrm{m0}}/3H_{0}^{2} = 0.25, 0.27, 0.30$
respectively ($H_{0} = 70\ \mathrm{km/s/Mpc}$). Dashed curves are
the spectra for the $\Lambda\mathrm{CDM}$ model. $n = - 0.5$ is
adopted for illustration. We see that a smaller
$\kappa\rho_{\mathrm{m0}}/3H_{0}^{2}$ can bring the shape of the
spectrum closer to the $\Lambda\mathrm{CDM}$ one: in this case
data on large scale structure cannot place stringent constraints
on the model.} \label{fig:8}
\end{figure}

We can now make a comparison between the matter power spectra in
metric and Palatini $f(R)$ cosmologies. In the later, it has been
shown that any small deviation from GR will cause the small-scale
spectra to differ from the GR one in observationally unacceptable
ways, either blowing up or oscillating and being prevented from
growing \cite{Koivisto2006, Li2006a, Li2006b}. This is because in
the Palatini $f(R)$ gravity $F$ and $T$ satisfy an algebraic
equation $F=F(T)$ such that those terms involving
$\hat{\nabla}^{2}F$ can always be rewritten as $\propto
\hat{\nabla}^{2}T\sim \hat{\nabla}^{2}\rho $ and consequently a
term $\propto k^{2}\Delta _{\mathrm{c}}$ appears in the growth
equation which makes the growths on small scales strongly
scale-dependent. In the metric $f(R)$ gravity here, in contrast,
$\epsilon $ is determined by the perturbations in $T$ through a
differential equation Eq.~(\ref{eq:57}). As $k$ increases, the
value of the perturbation in $F(\epsilon )$ decreases, so that
$k^{2}\epsilon $ does not exceed $\mathcal{O}(\kappa
\mathcal{X}a^{2})$ and becomes $k$ independent, so its effect
cannot be as exotic as that arising in Palatini $f(R)$ theories.

Before leaving this analysis, we want to briefly discuss the
cosmological viability of the present model. A more rigorous
analysis involves carefully searching the multidimensional
parameter space (as in \cite{Li2006a}) and is beyond the scope of
the present work. Firstly, we have seen that the background
evolutions allowed by the model are rather close to the $\Lambda
\mathrm{CDM}$ paradigm with particular choices of $n,$ and so
could be consistent with the SNIa observations. The confrontation
between predictions and the data on the linear power spectrum is a
little more complicated. For the CMB spectrum, Figure~\ref{fig:5}
indicates that it is similar to the $\Lambda \mathrm{CDM}$
prediction at higher $l$'s and can also reduce the quadrupole
power and bring it closer to the measured values \cite{Song2006}
(We have also checked that this low-$l$ power reduction is a
general feature of the model and is insensitive to the values of $H_{0}$ and $\rho _{\mathrm{m}0}$%
). However, because of the limitation from cosmic variance, the influences
on the likelihood analysis are expected to be small. For the matter
spectrum, at small scales the power is significantly higher than for the $%
\Lambda \mathrm{CDM}$ case, yet the fact that it has similar shape
to the latter indicates the possibility that a different choice of
galaxy bias might be able to make the model's predictions
consistent with, for example, the Sloan Digital Sky Survey (SDSS)
observations. However, we notice that at wavenumber $k\sim 0.05\
h\mathrm{Mpc^{-1}}$ deviations from the $\Lambda \mathrm{CDM}$
shape start to be significant, which might potentially be in
conflict with the data; this could still be alleviated by reducing
the quantity $\kappa \rho _{\mathrm{m0}}/3H_{0}^{2}$, as can be
seen in Figure~\ref{fig:8}.

We conclude that constraints on this model from large-scale structure could
be much weaker than those on the Palatini $f(R)$ models \cite{Li2006a,
Li2006b}. This said, it is still interesting to obtain quantitative joint
constraints from background cosmology, CMB, and matter power spectra on the
model, and examine whether it can be made consistent with other cosmological
observations (such as considered in \cite{Song2006}) in the future.

\section{Discussion and Conclusions}
\label{sect:Conclusion}

To summarise: in this article we have considered the cosmology of a subclass
of metric $f(R)$ gravity theories, that are characterised by $f(R)=R+\mu
^{2(n+1)}/(-R)^{n}$, and have discussed both the flat background Friedmann
universe and its inhomogeneous perturbations. For the background evolution,
we address the problem in both the Jordan frame and the Einstein frame and
find the correspondences between them. Generally, the evolution is attracted
towards a saddle point in the phase space which has the characteristics of a
matter-dominated era. If the initial conditions are chosen appropriately,
the universe can stay in the vicinity of that point for a sufficiently long
period of CDM-dominated evolution to occur. Finally, it always passes this
saddle point and evolves to a stable de Sitter point. The cosmic expansion
begins to accelerate during this transition period and a cosmic history that
is consistent with observations is possible in this model, as expected from
\cite{Amendola2006b}. In the Einstein frame, the model reduces to one with a
scalar field conformally coupled to (non-relativistic) matter, and it shares
some of the properties of a chameleon field cosmologies. That cosmological
model results in a matter-dominated era followed by an accelerating era can
also be seen from the analysis in this frame. The $\Lambda \mathrm{CDM}$
limit of the model is also discussed.

We derive the covariant and gauge invariant perturbation equations
for $f(R)$ gravity theories in general, which we believe to be
rather convenient for numerical calculations, and applied them to
the present model through a modified CAMB code. The linear spectra
are obtained and discussed. The CMB power spectrum displays
reductions in the low-$l$ power, which arises from a weakening of
the ISW effect because of insufficient late-time decay of the
large-scale potentials, as shown in the plots of the evolution.
For the matter power spectrum, we find that on small scales the
matter density growth is nearly scale-independent, making the
shape of the spectrum at large $k$'s similar to that of the
$\Lambda \mathrm{CDM}$ spectrum. Comparisons with the CMB and
matter power spectra in Palatini $f(R)$ gravity theories are then
made, which account for the dramatic differences between these two
approaches to modify GR, despite the fact that they could have the
same gravitational action.

Possible comparisons with different observational data sets are
also briefly discussed. It is found that neither the data on
background evolution, CMB spectrum, nor those on the matter power
spectrum can be used to exclude the model. Their joint constraints
on the model, however, could be complicated and involve making a
numerical search of the parameter space, which is beyond the scope
of the present work. However, we can see that constraints on this
metric $f(R)$ model could be much weaker than those on the
Palatini $f(R)$ theories, because for the later (1) the CMB power
is largely amplified at low $l$'s and (2) the matter power
spectrum at small scales is strongly scale dependent, both
conflicting with observations on CMB anisotropies and large scale
structure.

The form of $f(R)$ adopted here is one of the few which could
produce a viable model for the entire cosmic history. In this
model the modifications to GR take effect only at late times.
Therefore, it is interesting to look for other models in which the
modifications are significant at different times, along the lines
of refs. \cite{Li2006b, Bean2006}. In \cite{Li2006b}, a model with
$f(R)=R+\lambda _{1}\exp (R/\lambda _{2})$ is considered in the
Palatini approach and its effects on the linear power spectra are
also discussed. In \cite{Bean2006}, this same model is
investigated within the metric approach; however, it actually
gives a $\phi$-matter-dominated era ($\phi\mathrm{MDE}$),
\cite{Amendola2006a, Amendola2000}, rather than the standard
matter-dominated era, and is not cosmologically viable. Thus, it
may still be meaningful to look for viable early-time $f(R)$
cosmologies in
the metric formalism. Another interesting issue to explore is whether the $%
f(R)$ gravity models might be more adapted to hot dark matter. These topics
will be investigated elsewhere.

\

\begin{acknowledgments}
BL is supported by the Overseas Research Studentship, Cambridge
Overseas Trust and the Department of Applied Mathematics and
Theoretical Physics at the University of Cambridge. We are
grateful to Professor Ming-Chung Chu for early stimulating
discussions on this topic and to Professor Sergei Odintsov for his
helpful suggestions on the manuscript and pointing out references
to us. Also we would like acknowledge the various conversations
with Dr.~Van Acoleyen about the chameleon effect and solar system
constraints of the $f(R)$ model.
\end{acknowledgments}

\emph{Notes Added}: there is recently another paper
\cite{Faulkner2006} discussing the Chameleon properties of the
$f(R)$ theories, but the emphasis is different from ours here. See
also \cite{Navarro2006}.

\appendix



\begin{thebibliography}
\bibitem[Comment(2006)]{Comment1} It is worth noticing that the
evolution of the potential is much more complicated than that of a
simple weak decay. In fact, from Figure~\ref{fig:6} we can see
that for smaller scales the potential first grows and then decays.
These (rather rapid) growths of potential $\phi $ at later times
could also \emph{enhance} the ISW effect, which is determined by
the magnitude of
\[
I_{l}^{\mathrm{ISW}}=2\int^{\tau _{0}}\phi _{k}^{\prime
}j_{l}[k(\tau _{0}-\tau )]d\tau
\]%
where $j_{l}(k\tau )$ are the spherical Bessel functions and $\tau
_{0}$ the conformal time at present. This effect is more
significant for larger $|n|$s as can be seen in
Figure~\ref{fig:6}.
\end{thebibliography}
\end{document}